\documentclass[article,preprint,amssymb]{revtex4}

\usepackage{graphicx}

\begin{document}

\title{Fractional Charge and Quantized Current in the Quantum Spin Hall State}

\author
{Xiao-Liang Qi, Taylor L. Hughes, and Shou-Cheng Zhang}
\affiliation{Department of Physics, McCullough Building, Stanford
University, Stanford, CA 94305-4045\\
{\rm To whom correspondence should be addressed:}\\
{\rm sczhang@stanford.edu}}

\date{\today}

\begin{abstract}
A profound manifestation of  topologically non-trivial states of
matter is the occurrence of fractionally charged elementary
excitations. The quantum spin Hall insulator state is a
fundamentally novel quantum state of matter that exists at zero
external magnetic field. In this work, we show that a magnetic
domain wall at the edge of the quantum spin Hall insulator carries
one half of the unit of electron charge, and we propose an
experiment to directly measure this fractional charge on an
individual basis. We also show that as an additional consequence, a
rotating magnetic field can induce a quantized dc electric current,
and vice versa.
\end{abstract}

\maketitle

Soon after the theoretical proposal of the intrinsic spin Hall
effect\cite{murakami2003,sinova2004} in doped semiconductors, the
concept of a time-reversal invariant spin Hall
insulator\cite{murakami2004a} was introduced. In the extreme quantum
limit, a \emph{quantum} spin Hall (QSH) insulator state has been
proposed for various
systems\cite{kane2005,bernevig2006A,bernevig2006d}. The QSH
insulators are time-reversal invariant and have a bulk
charge-excitation gap. However, this system also possesses
topologically protected gapless edge states that lie inside the bulk
insulating gap. The edge states of the QSH insulator state differ
from the quantum Hall effect and have a distinct helical property:
two states with opposite spin-polarization counter-propagate at a
given edge\cite{kane2005,wu2006,xu2006}. The edge states come in
Kramers' doublets, and time reversal symmetry ensures the crossing
of their energy levels at special points in the Brillouin zone.
Because of this level crossing, the spectrum of a QSH insulator
cannot be adiabatically deformed into a topologically trivial
insulating state; therefore, in this precise sense, the QSH
insulators represent a topologically distinct new state of matter.

Recently, the QSH effect has been theoretically
proposed\cite{bernevig2006d} and experimentally
observed\cite{koenig2007} in HgTe quantum wells. In this experiment,
an applied gate voltage can tune the carrier type from n-type doping
to p-type doping, passing through a nominally insulating state. A
residual charge conductance approaching $2 e^2/h$ has been measured
in this insulating regime. Furthermore, the residual charge
conductance is independent of the width of the sample, indicating
that it is due to the helical edge state channels of the QSH
insulator.

Given the exciting theoretical and experimental development in this
field, one central question remains unanswered - what is the direct
experimental manifestation of this topologically non-trivial state
of matter? In the case of the quantum Hall (QH) effect, it is the
quantization of the Hall conductance and the fractional charge of
the elementary excitations which are a result of non-trivial
topological structure. The $Z_2$ topological invariant gives a
correct mathematical characterization of the QSH
state\cite{kane2005A}; however, unlike the TKNN quantum
numbers\cite{thouless1982} of the quantum Hall (QH) state, it is not
directly measurable experimentally. In this work, we show that for
the QSH state a magnetic domain wall induces an elementary
excitation with half the charge of an electron. We also show that a
rotating magnetic field can induce a quantized dc electric current,
and vice versa. Both of these physical phenomena are direct and
experimentally observable consequences of the non-trivial topology
of the QSH state. The idea of fractional charge induced at a domain
wall goes back to the Su-Schrieffer-Heeger (SSH) model\cite{su1979}.
For spinless fermions, a charge domain wall induces an elementary
excitation with one-half charge. However, for a real material such
as polyacetylene, two spin orientations are present for each
electron, and because of this doubling, a domain wall in
polyacetylene only carries integer charge; the beautiful proposal of
SSH, and its counter-part in field theory, the Jackiw-Rebbi
model\cite{jackiw1976}, have never been experimentally realized.
Conventional one dimensional electronic systems have four basic
degrees of freedom, \emph{i.e.} right and left movers with each spin
orientation. However, a helical liquid at a given edge of the QSH
insulator has only two: a spin up (or down) right mover and a spin
down (or up) left mover. Therefore, the helical liquid has half the
degrees of freedom of a conventional one-dimensional system, and
thus avoids the doubling problem. Because of this fundamental
topological property of the helical liquid, a domain wall carries
charge $e/2$. We propose a Coulomb blockade experiment to observe
this fractional charge. As a temporal analog of the fractional
charge effect, the pumping of a quantized charge current during each
periodic rotation of a magnetic field is also proposed. This
provides a direct realization of Thouless's topological
pumping\cite{thouless1983}.


\paragraph*{Theoretical description.} In the absence of time-reversal symmetry (TRS) breaking we can
express the effective theory for the edge states of a non-trivial
QSH insulator as
\begin{equation}
H_0=v_F \int dx (\psi^{\dagger}_{R+}i\partial_x
\psi_{R+}-\psi^{\dagger}_{L-}i\partial_x\psi_{L-})=v_F\int dx
\Psi^{\dagger}i\sigma^3\partial_x\Psi
\end{equation}\noindent where $\pm$ indicate members of a Kramers's doublet,
$L/R$ indicate left- or right-movers, and
$\Psi=(\psi_{R+},\psi_{L-})^T$\cite{wu2006,xu2006}. Note that these
helical fermion states only have two degrees of freedom since the
spin-polarization is correlated with the direction of motion.

Since the three Pauli spin matrices $\sigma_{1,2,3}$ are odd under
time-reversal, a mass term, being proportional to the Pauli
matrices, can only be introduced in the Hamiltonian by coupling to a
T-breaking external field such as a magnetic field or aligned
magnetic impurities.  To the leading order in perturbation theory, a
magnetic field generates the mass terms
\begin{equation}
H_M=\int dx \Psi^{\dagger}\sum_{a=1,2,3}m_a(x,t)\sigma_a\Psi=\int dx
\Psi^{\dagger}\sum_{a,i}{t}_{ai}B_i(x,t)\sigma_a\Psi\label{HM}
\end{equation}\noindent
where the model dependent coefficient matrix $t_{ai}$ is determined
by the coupling of the edge states to the magnetic field. According
to the work of Goldstone and Wilczek\cite{goldstone1981}, at zero
temperature the ground-state charge density and current in a
background field $m_a(x,t)$ is given by
\begin{eqnarray}
j_\mu=\frac1{2\pi}\frac{1}{\sqrt{m_\alpha
m^\alpha}}\epsilon^{\mu\nu}\epsilon^{\alpha\beta}m_\alpha\partial_\nu
m_\beta,~\alpha,\beta=1,2.\nonumber
\end{eqnarray}
with $\mu,\nu=0,1$ corresponding to the time and space components,
respectively. Note that $m_3$ does not enter the long-wavelength
charge-response equation. If we parameterize
$m_1=m\cos\theta,~m_2=m\sin\theta$, then the response equation is
simplified to
\begin{equation}
\rho=\frac{1}{2\pi}\partial_{x}\theta(x,t),\;\;\;
j=-\frac{1}{2\pi}\partial_{t}\theta(x,t).\label{GWformula}
\end{equation}

Such a response is ``topological" in the sense that the net charge
$Q$ in a region $[x_1,x_2]$ at time $t$ depends only on the boundary
values of  $\theta(x,t)$ \emph{i.e.}
$Q=\left[\theta(x_2,t)-\theta(x_1,t)\right]/2\pi$. In particular, a
half-charge $\pm e/2$ is carried by an anti-phase domain wall of
$\theta$, as shown in Fig. 1A\cite{jackiw1976}. Similarly, the
charge pumped by a purely time-dependent $\theta(t)$ field in a time
interval $[t_1,t_2]$ is  $\Delta Q_{\rm
pump}|_{t_1}^{t_2}=\left[\theta(t_2)-\theta(t_1)\right]/2\pi$. When
$\theta$ is rotated from $0$ to $2\pi$ adiabatically, a quantized
charge $e$ is pumped through the 1d system, as shown in Fig. 1B.

From the linear relation $m_a=t_{ai}B_i$, the angle $\theta$ can be
determined for a given magnetic field $B_i$ as
$\theta(x,t)=\theta({\bf B}(x,t))={\rm Im}\log\left({\bf
t}_1\cdot{\bf B}(x,t)+i{\bf t}_2\cdot {\bf B}(x,t)\right)$, in which
${\bf t}_{1(2)}$ is the 3d vector with  components $t_{1(2)i}$,
respectively. Since $\theta({\bf B})=\theta(-{\bf B})+\pi$, the
charge localized on an anti-phase magnetic domain wall of
magnetization field is always $\pm e/2$. For the pumping effect, the
winding number of $\theta(t)$ is given by the winding number of the
${\bf B}$-vector around the axis ${\bf t}_1\times {\bf t}_2$. The
conditions for these effects to be observed are
$k_BT,~\hbar\omega\ll E_g$, where $\omega$ is the pumping frequency,
and $E_g=\sqrt{({\bf t}_1\cdot {\bf B})^2+({\bf t}_2\cdot {\bf
B})^2}$ is the energy gap of the helical edge state generated by the
magnetic field.

After the general analysis, we now discuss more details of the
experimental realization of such topological effects. First, to gain
intuition about the energy scales in this problem, we consider the
coefficients $t_{ai}$ for ${\rm HgTe/CdTe}$ quantum wells which can
be obtained numerically. By solving the four band effective model
given in Ref. \cite{bernevig2006d} with an open boundary we obtain
the wavefunctions of the two edge states $\left|k;\pm\right\rangle$.
The effective $2\times 2$ Hamiltonian (\ref{HM}) of the edge states
including magnetic field effects is obtained by standard
perturbation theory, from which the coefficients $t_{ai}$ are
extracted. For a quantum well with thickness $d=70\AA$ and an edge
along the $y$-direction, we obtain ${\bf
t}_1=\left(-0.3,0,0\right)~{\rm meV/T}$ and ${\bf
t}_2=\left(0,-0.3,-3.1\right)~{\rm meV/T}$.\cite{liu2007,novik2005}
(Here and below the $z$ direction is the quantum well growth
direction.) Thus, the gap induced by an in-plane field $B_x=1{\rm
T}$ is $E_{gx}\simeq 0.3{\rm meV}$, while the perpendicular
component $B_z=1{\rm T}$ produces a much larger gap $E_{gz}\simeq
3.1{\rm meV}$. (Such a large anisotropy between in-plane and
perpendicular magnetic field agrees well with the experimental
observations in Ref.\cite{koenig2007}.) Consequently, the charge
fractionalization effect can be observed at temperature $T\ll 35{\rm
K}$ on a domain wall with a {\em perpendicular} magnetic field,
while the adiabatic pumping effect can only be observed at much
lower temperatures $T\ll 3.5{\rm K}$ since it depends partially on
the gap from in-plane fields.
%

We have shown through the discussions above that the fractional
charge and adiabatic pumping effects are experimentally feasible. In
the following paragraphs we propose detailed experimental settings
needed to observe these two topological effects.

\paragraph*{Observation of fractional charge on the domain
wall.} Recently, a novel device has been developed to measure the
charge of a confined region: the single-electron transistor
(SET)\cite{kastner1987,kastner1992,yoo1997}. When applying a gate
voltage $V$ on top of a confined region (e.g., quantum dot or wire)
with capacitance $C$, the Coulomb energy is given by $E_c
(Q)=Q^2/2C+VQ$. The number of charges trapped in the confined region
is quantized as $Q=Ne$, with the integer $N$ determined by
minimizing the energy $E_c(Ne)=\min_{n\in\mathbf{Z}}E_c(ne)$. (Here
and below ${\bf Z}$ stands for the set of all integers.) Thus an
additional electron entering the confined region will cost an energy
$\Delta E=E_c((N+1)e)-E_c(Ne)$. A straightforward calculation shows
that $\Delta E=e^2/2C$ for $CV/e\in \mathbf{Z}$, while $\Delta E=0$
for $CV/e-1/2\in \mathbf{Z}$. Consequently, the two-terminal
conductance of the device is governed by an activation behavior
$G=G(V)\propto e^{-\Delta E/k_BT}$ and shows oscillations with
period $\Delta V=e/C$ and peak positions at
$V=\left(n+1/2\right)e/C,~n\in\mathbf{Z}$. In other words, the
conductance peak appears when the net charge of the confined region
is a half-odd integer times the electron charge. This allows one to
sensitively measure charges comparable to or even smaller than the
electronic charge\cite{yoo1997,martin2004}.

The fractional charge created by a magnetic domain wall on the QSH
edge is confined in the region between the two magnetic domains
separated by the wall. This confined charge can be measured by
designing a magnetic SET experiment. A schematic picture of such a
device is shown in Fig. 2. Two magnetic islands can trap the
electrons between them, just like a quantum wire trapped between two
potential barriers\cite{kastner1992}. In such a device, the
conductance oscillations can be observed as in usual Coulomb
blockade measurements. The background charge in the confined region
consists of two parts: $Q_b=Q_c+Q_e$, with $Q_e$ the contribution of
the lowest subband electrons and $Q_c$ that of higher energy bands
and nuclei. When the field direction in one of the magnetic domains
is switched, $Q_c$ remains invariant but $Q_e$ will change by $e/2$
which demonstrates the half-charge associated to the antiphase
domain wall. Consequently, if we use a top gate on the confined
region and measure the conductance oscillations $G(V)$, there will
be a half-period phase shift between the oscillation pattern of
parallel and anti-parallel magnetic domains, as shown in Fig. 2.

Experimentally,  magnetic islands can be deposited on top of
semiconductor heterostructures creating a hybrid
ferromagnet-semiconductor device\cite{prinz1990,halm2007}. The
magnetic islands can be polarized via magnetic field and locally
switched using a coercive field and conventional magnetic-force
microscopy techniques (see \emph{e.g.} \cite{kleiber1998}). The
quantum well will be locally exposed to the fringe fields of the
ferromagnetic islands. To observe the conductance oscillations,
several conditions should be satisfied by the magnetic field
configuration:
\begin{enumerate}
\item The oscillation period $\Delta V=e/C$ should be much smaller
than the bulk gap scale $V=E_g/e$ so that several periods of
oscillation can be observed before the gate voltage is so high that
the bulk states are activated. This leads to the requirement $C\gg
e^2/E_g$.\\
\item The ratio of the minimal conductance to the maximal
conductance is estimated by
\begin{eqnarray}
G_{\rm min}/G_{\rm max}\simeq
\exp\left[-\frac{e^2/2C}{k_BT}\right]\nonumber
\end{eqnarray}
with $e^2/2C$ the maximal charge activation gap. For the
oscillations to be observable this ratio should be reasonably
smaller than one, which leads to the condition
$e^2/2C\geq k_BT$ or $C\leq e^2/2k_BT$.\\
\item The domain wall state trapped between the magnetic domains
has an exponential tail as shown in Fig. 2. To make the conductance
measurement, the size of the magnetic islands should be comparable
to the exponential tail length $\xi_M\simeq \hbar v_F/E_g$ with
$E_g$ the magnetic field-induced gap, so that the localized state is
well-confined, but tunnelling through the barrier is still strong
enough to support observable transport.
\end{enumerate}

The edge state trapped between magnetic domains has the linear size
$\xi\times d\times L$ where $\xi\simeq \hbar v_F/M$ is the
penetration depth of the edge state, $d$ is the thickness of quantum
well, and $L$ is the distance between the two magnetic regions.
Under the one-dimensional approximation $L\gg d,~\xi$ we obtain the
approximate form of the capacitance  $C\simeq 4\pi\epsilon_0
L/\left[\log \left(L^2/d\xi\right)-2\right]$. For $d=70\AA$ we find
the condition $1{\rm \mu m}\ll L\ll 100{\rm \mu m}$. For a magnetic
field $B_z=1T$ the size of each magnetic island is $r\sim 120 $ nm.

So far in our discussions, the magnetic domain wall structure is
externally imposed. However, it is also possible that the system can
spontaneously generate such  magnetic domain walls. In Ref.
\cite{wu2006,xu2006}, it was shown that two-particle backscattering
interactions are allowed in the helical liquid. In the strong
coupling limit, such a process can lead to a spontaneous breaking of
TRS, and  spontaneous generation of a magnetic moment at the edge.
This symmetry breaking is described by an Ising like $Z_2$ order
parameter, which at any finite temperature in 1d leads to a finite
density of magnetic domain walls. Our work shows that such domain
walls will carry fractional charge $\pm e/2$, which are the
elementary excitations of the system.

We remark that the phenomenon of  fractional charge associated with
a magnetic domain wall is an example of ``electro-magnetic duality"
in 1d. Here, a domain wall is a point-like object, and is dual to a
point particle. In our particular case, a {\it magnetic domain wall}
induces an {\it electric point charge} $\frac{e}{2\pi}\Delta\theta$.
In 3d, a natural magnetic point singularity is a magnetic monopole,
and Witten showed\cite{witten1979} that it can induce an electric
point charge $\frac{e}{2\pi}\theta$, where $\theta$ is the vacuum
angle of quantum-chromodynamics. The duality between a magnetic
point charge and an electric point charge in the helical liquid of
the QSH state has many other profound consequences which we shall
demonstrate in future publications.

\paragraph*{Observation of the quantized charge current.} If the magnetic moment of one
domain in the proposed SET device is rotated continuously by a full
period while the other one remains static, the conductance peak
position will shift relatively by a full period, as shown in Fig.
2B. Such a shift of the conductance peaks shows the change of
background charge by $e$ in the confined region, which is a
consequence of the topological pumping effect. Another device to
measure the pumping current directly is shown in Fig. 3. One
strongly-pinned and one easy-plane magnetic island are deposited
above one of each of the two arms of the device, respectively. When
applying a small rotating external field with frequency $\omega$,
the magnetization of the easy-plane island will be rotated while the
pinned one remains static. Consequently, a quantized charge current
is pumped, given by the formula $I=e\omega/2\pi$ under the adiabatic
approximation. To observe such an effect the magnetic field-induced
gap must be much larger than the temperature, which will require
${\rm Mn}$ doping as discussed earlier.

\paragraph*{Conclusion and more discussions.} In conclusion, based on both
the general theoretical discussions and quantitative numerical
calculations, we have shown the feasibility of two striking
topological effects in the QSH state. Using a single-electron
transistor-like sensor we proposed an experimental setting to create
and observe the fractionally charged domain wall. Such topological
phenomena, if observed, not only provide the first experimental
realization of the fractional charge in one-dimensional systems, but
also introduce a \emph{physical and operational} definition of the
two-dimensional topological (QSH) insulator. An (infinitesimal)
magnetic field
 domain wall configuration can be used as a sensor
to characterize  two dimensional insulators. If such a detection
device induces a localized fractional charge response on the sample
edge, then the system is defined to be a topological insulator. Such
a definition is experimentally meaningful since it is based on the
response of the system to some physical external field, and is
completely analogous to the definition of the quantum Hall insulator
as system which produces a quantized Hall response to an external
electric field.

{\bf Acknowledgement.}---We wish to thank B. A. Bernevig, H.
Buhmann, X. Dai, M. Koenig, C.-X. Liu, L. Molenkamp for insightful
discussions. This work is supported by the NSF through the grants
DMR-0342832, by the US Department of Energy, Office of Basic Energy
Sciences under contract DE-AC03-76SF00515, and Focus Center Research
Program (FCRP) Center on Functional Engineered Nanoarchitectonics
(FENA).

\newpage

\begin{figure}[h]
\begin{center}
\includegraphics[width=6in] {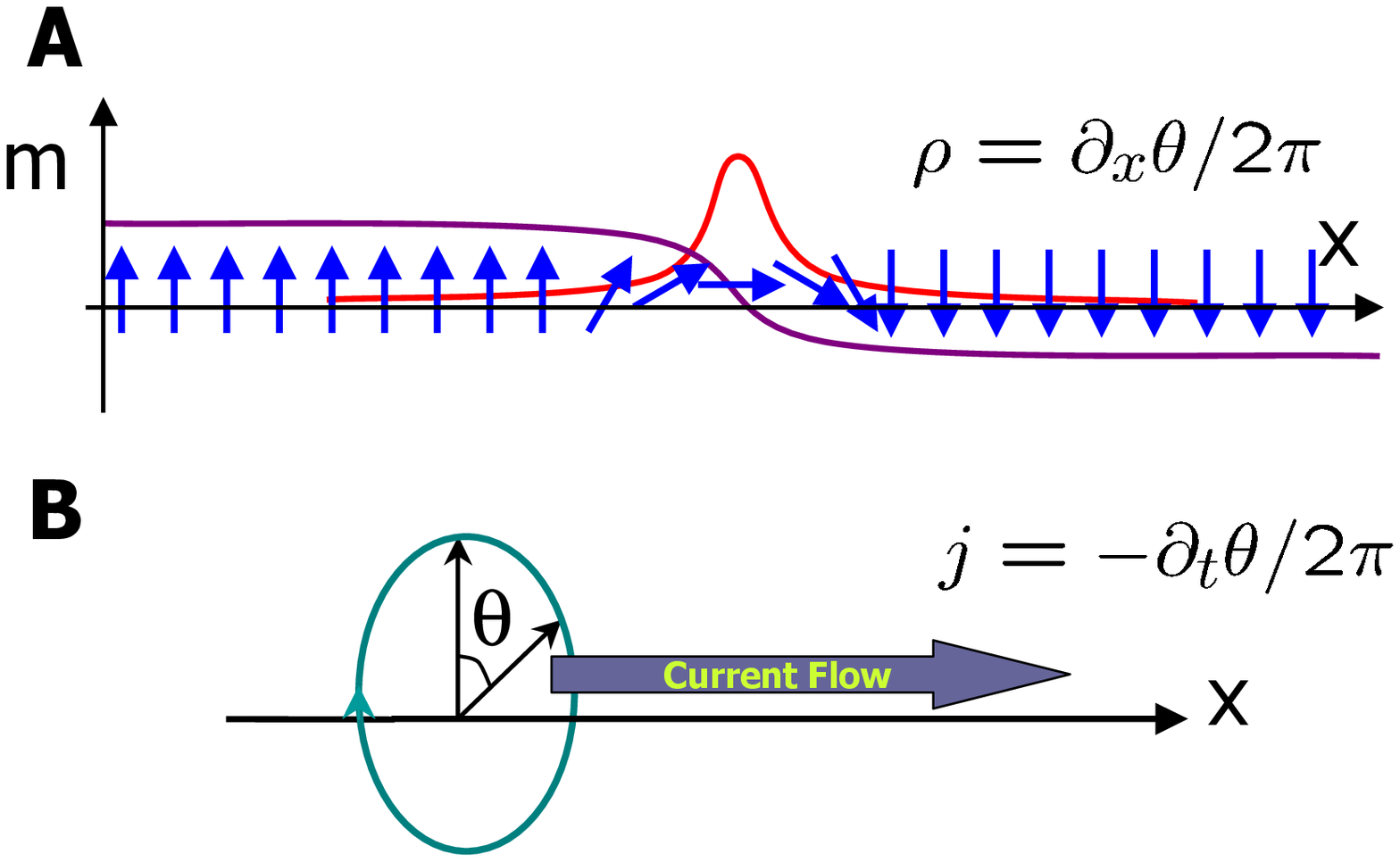}
\end{center}
\caption{{\bf A}. Schematic picture of the half-charge domain wall.
The blue arrows show a magnetic domain wall configuration and the
purple line shows the mass kink. The red curve shows the charge
density distribution. {\bf B}. Schematic picture of the pumping
induced by the rotation of magnetic field. The blue circle with
arrow shows the magnetic field rotation trajectory.}
\end{figure}

\newpage

\begin{figure}[h]
\begin{center}
\includegraphics[width=6in] {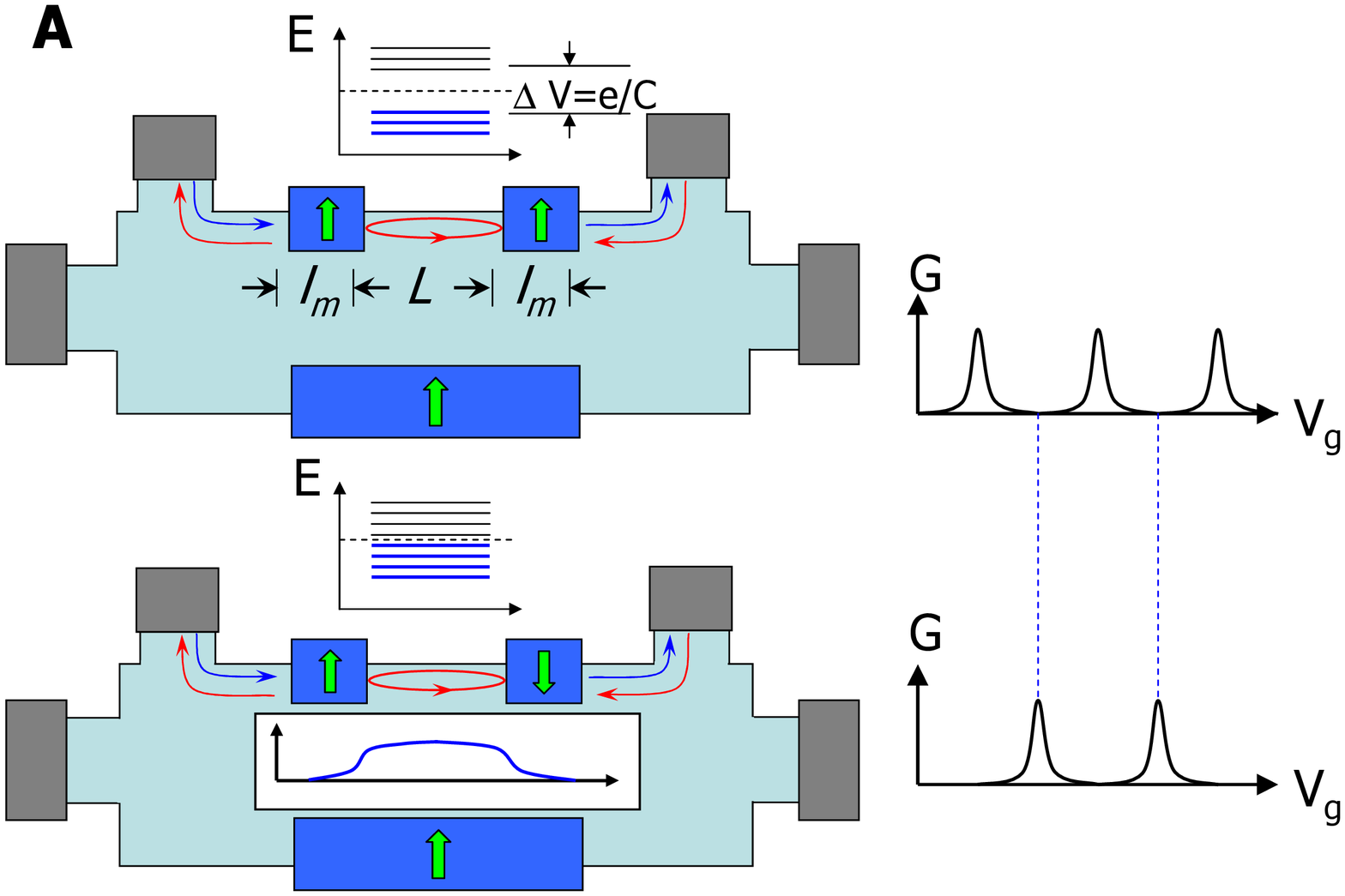}\\
\includegraphics[width=6in] {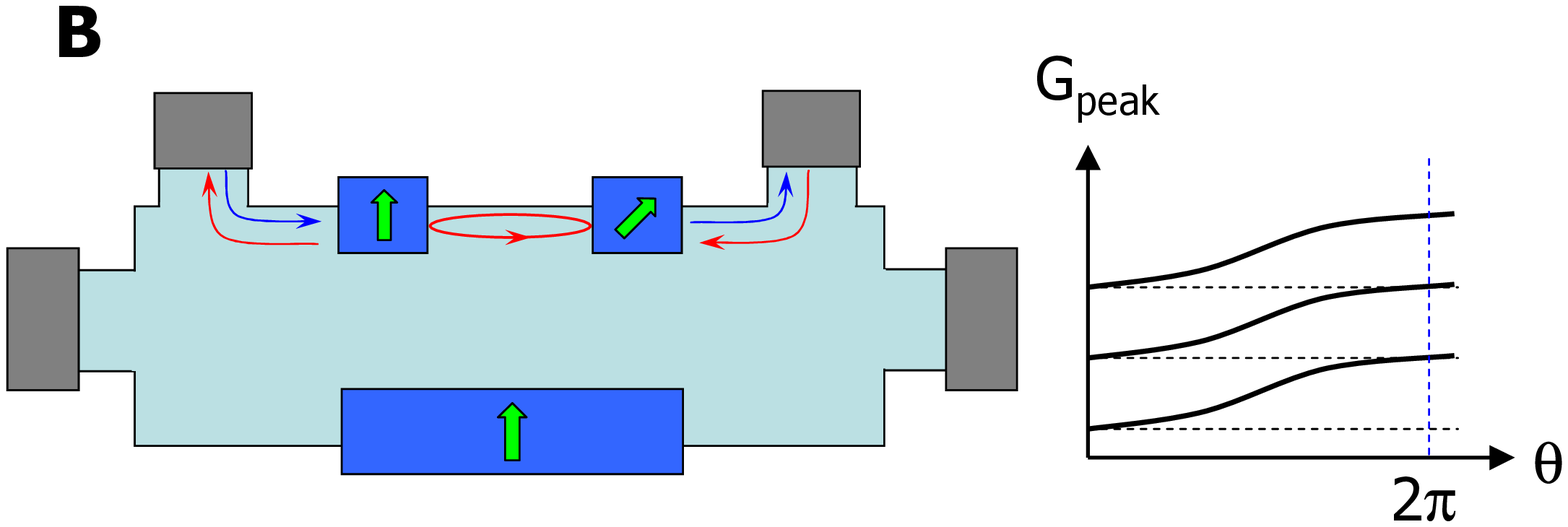}
\end{center}
\caption{{\bf A}. The SET device with parallel and anti-parallel
magnetic domains. The blue and grey rectangles are the magnetic
domains and voltage probes, respectively. (In all the figures, the
magnetic domain on the bottom arm is always pinned, which is
designed to block the transport of that arm so that the conductance
of the SET device on the upper arm can be measured.) The conductance
peak shift is shown on the right. The inset shows a schematic
picture of the bound state wavefunction. {\bf B}. Schematic picture
of the conductance peak positions $G_{\rm peak}$ being shifted by
continuous rotation of the magnetic field. (Here and in Fig. 3, the
actual direction of the rotating magnetic field should be in the
plane {\em perpendicular} to the edge.)}
\end{figure}

\newpage
\begin{figure}
\begin{center}
\includegraphics[width=4in] {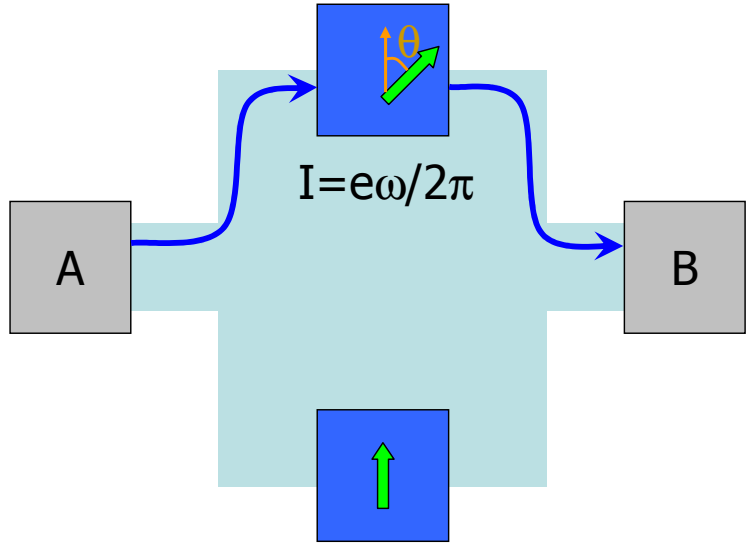}
\end{center}
\caption{Schematic picture of the device for measuring the quantized
charge current. A and B are source and drain without an applied
voltage bias between them.}
\end{figure}

\end{document}